# Portlandite solubility and $Ca^{2+}$ activity in presence of gluconate and hexitols

**BOUZOUAID Lina[1], LOTHENBACH Barbara[2], FERNANDEZ-MARTINEZ Alejandro[3], LABBEZ Christophe[1]**

[1] ICB, UMR 6303 CNRS, Univ. Bourgogne Franche-Comté, FR-21000 Dijon, France

[2] Empa, Concrete & Asphalt Laboratory, Dubendorf, Switzerland

[3] Univ. Grenoble Alpes, Univ. Savoie Mont Blanc, CNRS, IRD, IFSTTAR, ISTerre, 38000 Grenoble, France.

## Abstract

The current paper investigates the impact of gluconate, D-sorbitol, D-mannitol and D-galactitol on calcium speciation at high pH values by i) solubility measurements of portlandite ($Ca(OH)_2$) and ii) potentiometric titration measurements of calcium salt solutions. Thermodynamic modeling was used to fit the chemical activities of $Ca^{2+}$ and $OH^-$ ions and thus to determine the strength and kind of the different Ca-organic-hydroxide complexes. The strength of complex formation with Ca decreases in the order gluconate >> sorbitol > mannitol > galactitol, which follows the same order as sorption on portlandite. Heteropolynuclear gluconate complexes with calcium and hydroxide dominate the Ca-speciation in the presence of portlandite, while for sorbitol ternary $CaSorbOH^+$ complexes were dominant under alkaline conditions. We expect that these results will help in better understanding the influence of gluconate and hexitols on the hydration of alite and Portland cement.

## 1. Introduction

The chemical activity of ions, *a*, and the solubility of minerals are crucial factors in determining the thermodynamic conditions and the driving force of a mineral to dissolve or to precipitate (1). The extent to which a solution is out of equilibrium is given by the deviation from the theoretical solubility and is quantified by the saturation index. For a solid such as portlandite, of solubility product $K_{Ca(OH)_2}$, the saturation index (SI) is defined as:

$$SI = \log\left(a_{Ca^{2+}} \cdot a_{OH^-}^2 / K_{Ca(OH)_2}\right)$$

The knowledge of the elemental concentrations, the speciation and the ion activities provides a simple measure for the driving force of dissolution or precipitation reactions. However, activity of ions is sensitive to the presence of other chemical species via the effect of ionic strength. Many ions can also form different soluble complexes such as e.g. calcium gluconate complexes: $Cagluc^+$, $Ca(gluc)_2^0$, ... and can in addition promote or inhibit crystal growth/dissolution, which can make the determination of the activity of ions difficult and tedious.

In the case of concrete, organic molecule inhibitors, called retarders, are commonly used in specific applications (2) (3) (4) to delay the cement setting. Superplasticizers, often comb co-polymers, employed in the formulation of ultra high performance concrete, are also known to retard the curing of concrete (5) (6) (7). Although used in low amount, less than 1 wt.% of Portland cement (8), the concentration of organic molecules in the solution present in the interstitial pores formed by cement grains, easily reaches several tens of millimolar concentrations during the first hours after mixing with water and can thus impact the cement reactions involved in the curing. These effects have been illustrated in several studies with various organic molecules (9) (10). Invariably, it was found that the organic molecules acting as cement curing inhibitors can greatly influence the elemental concentrations in the aqueous pore solution and retard the cement hydration. The cement hydration is complex and dependent on two interrelated and concomitant processes, which are the dissolution of the cement grains and the precipitation of the hydrates. However, the physical and chemical mechanisms responsible for this retardation are not fully understood (11) (12) (13) (14) (15).

Furthermore, in those organic-cement systems, an accurate quantification of the dissolution rate of cement anhydrates and precipitation rate of cement hydrates at well-defined *SI* is still missing as a result of limited knowledge of complex formation.

For the organics of interest in this paper, the pure dissolution of alite ($Ca_3SiO_5$) was found to be negligibly affected by gluconate and three different hexitols, namely sorbitol, mannitol and galactitol (15). These organics, however, have a great impact on alite hydration and to a less extend on the hydration of Portland cement (15). These molecules were thus conjectured to mostly act as nucleation-growth inhibitors of calcium silicate hydrate, C-S-H (16) In a recent experimental study, it was suggested that the inhibition of the crystallization of portlandite, $Ca(OH)_2$, could be the main reason of the slowdown of alite hydration (17).

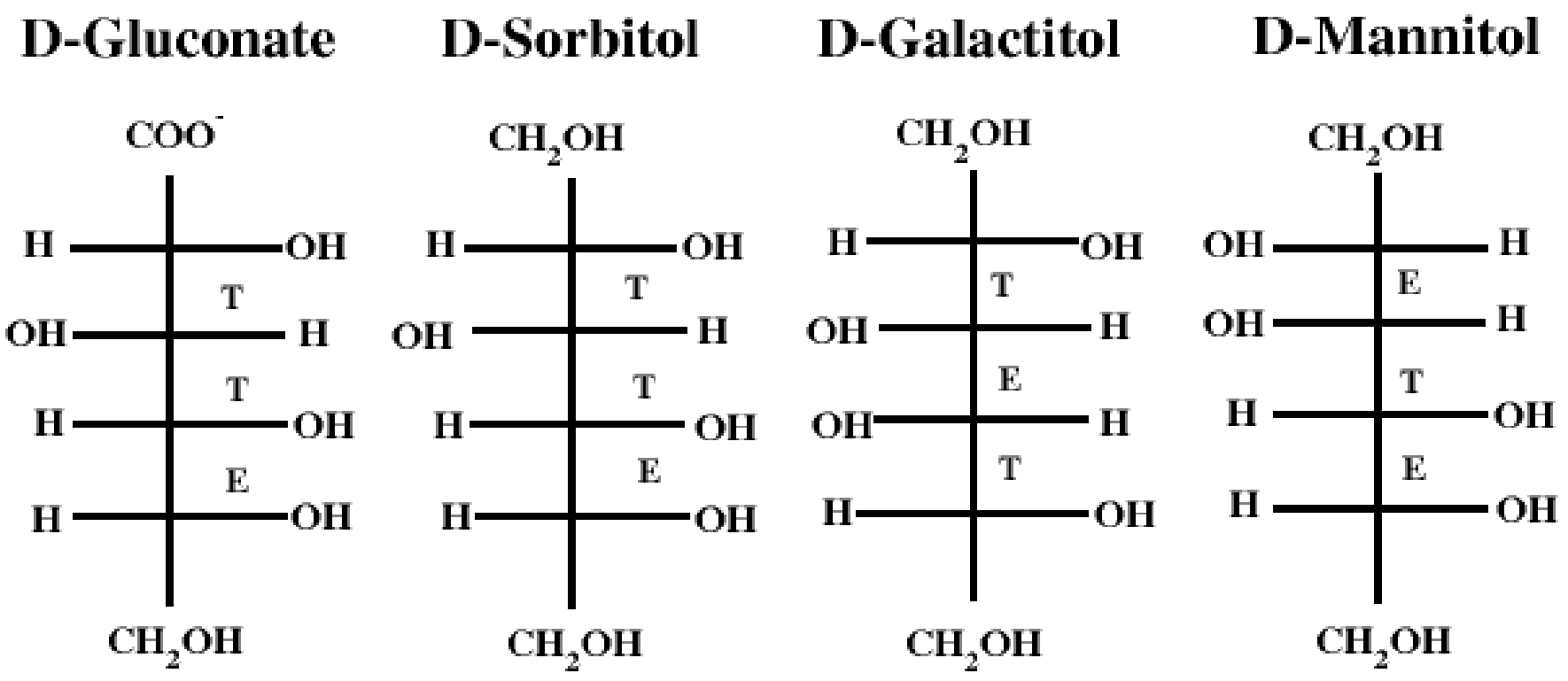

Figure 1: Structure of (from left to right) D-gluconate, D-sorbitol, D-galactitol and D-mannitol. ”T” corresponds to the threo diastereoisomer configuration. “E” corresponds to erythro diastereoisomer configuration.

Gluconate, a negatively charged molecule, was found to be a stronger retardant of cement hydration than neutral hexitols (16) (18) (19). For different hexitols also the stereochemical arrangements of the organic molecules, as illustrated in Figure 1, has an influence (16). The retardation was observed to increase from mannitol to galactitol to sorbitol, which have the same chemical composition and functional groups, but a different arrangement. However, the chemical mechanisms, which could explain how these organics retard are poorly understood, e.g. their impact on the anhydrate and hydrate solubility and the eventual formation of aqueous complexes [ (16) (18)]. In particular, in the high pH range little is known about possible aqueous complexes with organic molecules or their stability.

It has been suggested that the ability of the organic molecules to form complexes with $Ca^{2+}$ is directly correlated to their adsorption affinity with calcium rich surfaces of C-S-H and alite (18) (19) which, in turn, may impart the surface tension of the nucleus or the rate of attachment of species to the nucleus and, thus, the nucleation rate. Thus, the understanding and quantification of complex formation between organic molecules and calcium could be fundamental for a better understanding of the observed retardation by organic additives[1]. In addition, only an adequate quantification of the different Ca-complexes formed in the presence of organics makes it possible to determine the ion activities needed to calculate *SI* with respect to $Ca_3SiO_5$, calcium silicate hydrate and portlandite.

D-gluconate, is a well know retarding additive (20) (21) (22) widely used in the industry. In addition to its role as retarder in cements and concretes, gluconate is also used for water treatment and metal surface treatment due its strong complexation ability with cations. The complex formation between Ca and gluconate has been investigated in several studies (23) (24) (25) (26) (27) (28), but mainly at high ionic strength (1.0 M NaCl) and relatively low pH. On the other hand, the complex formation between hexitols and calcium ions was much less investigated (29) (30) and again at relatively low pH values not relevant for cements.

The present paper thus aims to investigate the speciation of alkaline calcium solutions in the presence of a carboxylate sugar acid: gluconate and several uncharged hexitols: D-sorbitol (D-glucitol), D-mannitol, and D-galactitol, at concentrations and pH values relevant for cementitious systems. A particular emphasis is on the ability of the organics to form complexes with calcium ions. This is assessed experimentally by solubility measurements of portlandite and ion activity measurements of alkaline calcium solutions in presence of increasing amount of organics. The results were then fitted with a speciation model, using the open source software PHREEQC, to determine the strength and the various types of calcium complexes with the organic molecules.

[1] Numerous more effects may impart a retardation of cement hydration

## 2. Materials and methods

*2.1 Materials*

The different stock solutions from each compound were prepared by dissolving $Ca(NO_3)_2$ $4H_2O$ (Sigma-Aldrich, ≥99% purity), potassium gluconate ($C_6H_{11}KO_7$, Sigma-Aldrich, ≥99% purity) D-sorbitol ($C_6H_{14}O_6$, Sigma-Aldrich, ≥99% purity), D-mannitol ($C_6H_{14}O_6$, Sigma-Aldrich, ≥99% purity), and D-galactitol ($C_6H_{14}O_6$,, Sigma-Aldrich, ≥99% purity), in boiled and degassed milliQ water. In the potentiometric titration experiments, potassium nitrate ($KNO_3$, Sigma-Aldrich, ≥99% purity) was used as a background electrolyte (0.1 M) and KOH ( >85%, Sigma-Aldrich) to increase the pH values to 11.3, 12.3, 12.7 and 13.0. For the solubility measurement, portlandite, calcium hydroxide ($Ca(OH)_2$, Sigma-Aldrich, ≥ 95%) was used.

It is important to note that the hexitols used in this study are all isomers, sharing the formula, $HOCH_2(CHOH)_4CH_2OH$, but differ in the stereochemical arrangement of the OH groups as illustrated in Figure 1.

*2.2 Solubility experiments*

For the solubility experiments, various series of samples were prepared in a glove box with 3.92 g of $Ca(OH)_2$, as a solid buffer, enclosed in a dialysis membrane and then placed in a 250 mL polypropylene flask filled with 200 mL of $CO_2$-free solution with different amounts and type of organic molecule, see Figure 2. Prior to use, the dialysis membranes (Spectra / Por, MWCO 12-14 kD) were dipped in distilled water for 30 minutes to remove any organic residues and dried in a desiccator overnight; the dialysis bags were closed with polyamide clamps (Carl Roth, length 50 mm). Finally, the samples were stored in a 16 L plastic barrel filled with $N_2$ gas, to guarantee $CO_2$ free conditions, and placed on a shaking table during four weeks at 23°C to ensure a proper equilibrium.

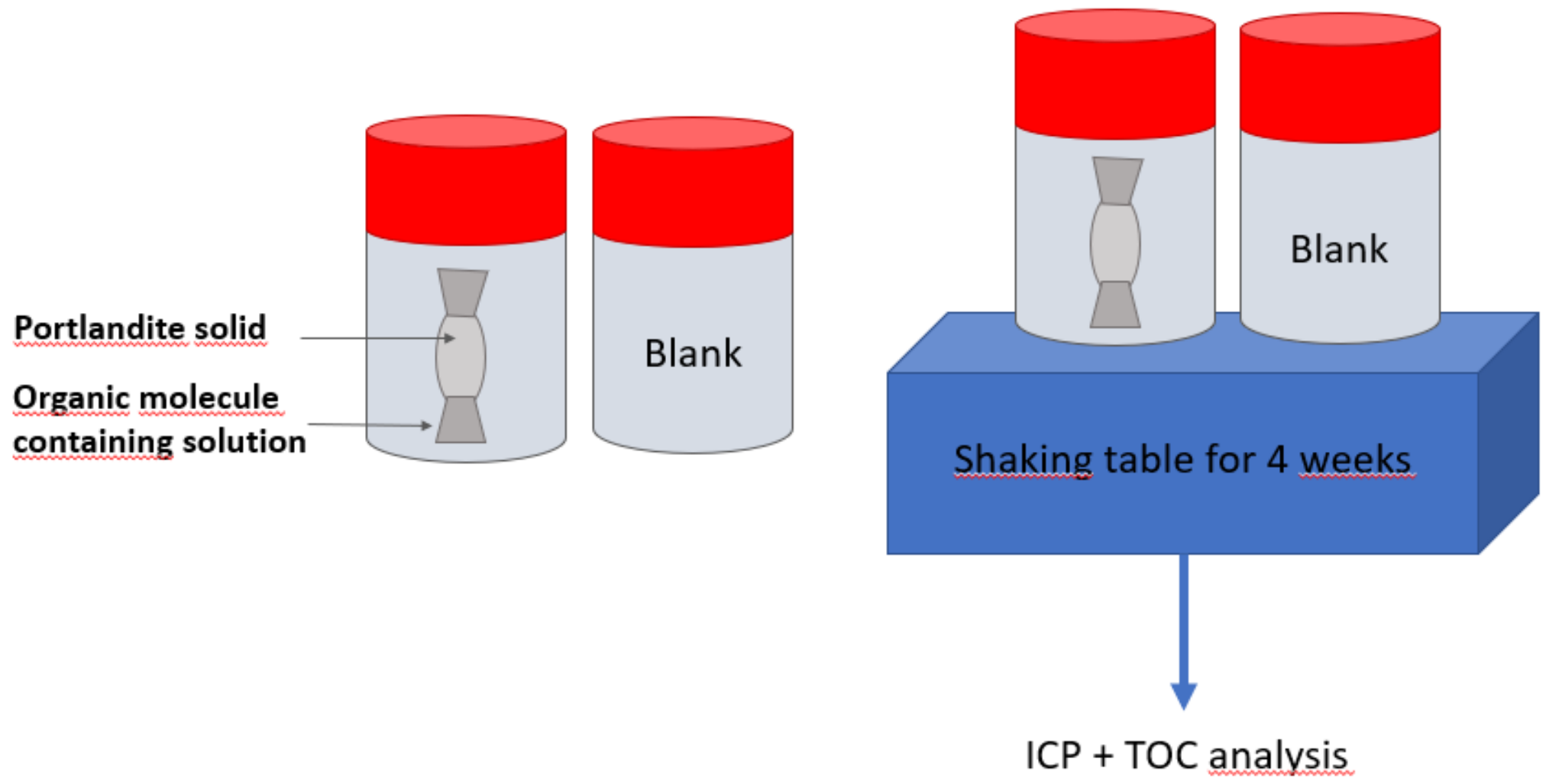

Figure 2: Schematic representation of a pair of samples used for the solubility and adsorption experiments. The flask used for the solubility measurement contains a dialysis bag filled with portlandite powder, immersed in a solution containing the organic molecule. The second flask contains the organic molecule solution only and is used as a blank/reference to verify the organic concentration introduced initially. This allows the determination of the organic adsorption on C-S-H by mass balance based on the measured difference. Note that the same stock solution is used for each sample pair.

The pH values were recorded after removing the dialysis bags from the bottles. The pH electrode (Consort C931 electrochemical analyser) was calibrated using Sigma Aldrich buffer ( pH 4, 7, 9 and 12). The total concentration of the elements Ca and Si was measured by inductively coupled plasma-optic emission spectroscopy (ICP-OES 5110, Agilent). The bulk concentration of organics at equilibrium was measured as total organic content with a TOC VCPN instrument (Shimadzu). This method is based on the oxidation of organic molecules contained in solution by gaseous oxygen with a platinum-based catalyst in an oven raised to a temperature of 720°C. The $CO_2$ formed is detected by Infra-Red Non-Dispersive (NDIR). The detection threshold for this device is very low (4µg/L).

*2.3 Titration of $Ca^{2+}$ with the ion selective electrode*

The chemical activity of calcium ions was also determined from potentiometric titration measurements using an automatic titrator instrument (Metrohm 905 Titrando); the setup is detailed in Figure 3. All measurements were performed in a titration reactor thermostated to 23.0 ± 0.1°C.

The titrated solutions were continuously stirred at a constant rate 430 rpm. A nitrogen flow circulated continuously above the solution to avoid the ingress of $CO_2$. It was taken care that the gas did not enter the solution to avoid any disturbance of the electrodes. In the reactor, 100 mL of a solution containing 0.1 M $KNO_3$ background electrolyte and 0.25 mM $Ca(NO_3)_2$ was thermostated for approximately 20 minutes before the start of the titration. After 3 minutes of equilibration time, and only if the change in the electrode signal was < 0.5 mV/min, a titrant solution containing 0.2M of organic molecule was added drop by drop (0.2 ml, 50 times for a total volume added of 10 ml) at the maximal speed registered in the software "Tiamo" (Metrohm).

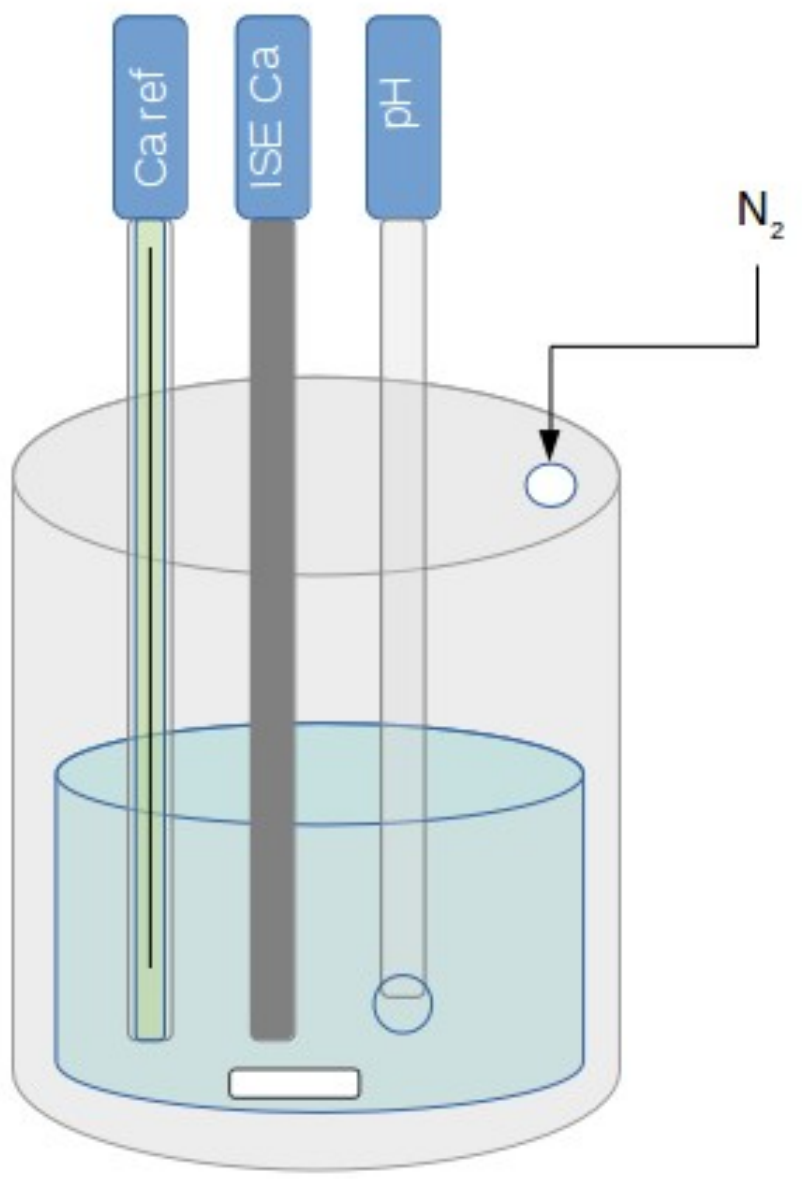

Figure 3: Schematic representation of the experimental titration set up: the reactor contains the titrated solution, the calcium specific electrode, the reference electrode, and the pH electrode.

The $Ca^{2+}$ activity was measured at 23.0 ± 0.1°C with a calcium sensitive electrode (Metrohm Ca ISE, 6.0508.110) coupled to a reference electrode (Metrohm Ag, AgCl/3 M KCl, 6.0750.100). A stable electrode signal could only be obtained with the use of a background electrolyte. The titrations were thus performed in 0.1 M $KNO_3$ to ensure a stable signal and to limit the influence of the background electrolyte on the complex formation. We have chosen potassium nitrate as $K^+$ interferes less with the $Ca^{2+}$- selective electrode than $Na^+$ (31). The $Ca^{2+}$ electrode was calibrated prior to the measurements by a titration of 100 mM $KNO_3$ solution with a solution containing 500 mM $Ca(NO_3)_2$ (0.2 ml, 50 times for a total volume added of 10 ml) and by plotting the measured mV against the calculated $Ca^{2+}$ activity calculated with PHREEQC as detailed below. The response of the $Ca^{2+}$ electrode was found to be linear with a slope of 29±1 mV, which corresponds to the expected slope of 29.4 mV at 23°C. The pH was determined with a pH electrode (Metrohm pH Unitrode with Pt 1000, 6.0259.100), which allows reliable measurements up to pH = 14. The pH electrode was calibrated prior to the measurements with standard buffer solutions (pH 9, 10, and 12.45 from Sigma Aldrich).

### *2.4 Thermodynamic simulation and complexation constants*

**Table 1.** Complex formation constants K, expressed in log K, between calcium and gluconate ($Gluc^-$) at standard conditions (1 bar, 25°C) reported in literature and determined in the present study.

| | Sayer (28) 0M | Masone (27) *(0.5M)* 0M[1] | Zhang (35) *(0.1M)* 0M | Pallagi (24) *(1M)* 0M | Bretti (36) 0M | Kutus (26) *(1M)* 0M | This study 0 M |
|---|---|---|---|---|---|---|---|
| Solid | | | | | | | |
| $Ca^{2+} + 2OH^- = Ca(OH)_2$ | - | - | - | - | - | - | **-5.20** [a] |
| Aqueous complexes | | | | | | | |
| $Ca^{2+} + OH^- = CaOH^+$ | - | - | - | *(0.97)* 1.83 | - | - | **1.22** [a] |
| $GlucH^0 = Gluc^- + H^+$ | 3.7 | - | *(3.30)*3.53 | *(3.24)* 3.64 [b] | 3.71 | - | **3.64** |
| $Gluc^- + OH^-$ $= GlucOH^{2-}$ [c] | - | - | - | *(0.08)* -0.44 | - | - | **-0.44** |
| $Ca^{2+} + Gluc^-$ $= CaGluc^+$ | 1.21 | *(1.05)* 1.79 | - | *(0.37)* 1.23 | - | *(0.70)* 1.56 | **1.56** |
| $Ca^{2+} + 2Gluc^-$ $= CaGluc_2^0$ | - | *(1.88)* 2.98 | - | - | - | *(1.65 [d])* 2.9 | **2.85** |
| $Ca^{2+} + OH^- + Gluc^-$ $= CaGlucOH^0$ | - | - | - | *(2.82)* 4.07 | - | *(2.86 [d])*4.11 | **3.95** [e] |
| $2Ca^{2+} + 3OH^- + Gluc^-$ $= Ca_2Gluc(OH)_3^0$ | - | - | - | *(8.04)* 10.48 | | - | **-** |
| $2Ca^{2+} + 4OH^- + 2Gluc^-$ $= Ca_2Gluc_2(OH)_4^{2-}$ | - | - | - | - | - | *(9.49 [d])* 11.34 | **11.25** [e] |
| $3Ca^{2+} + 4OH^- + 2Gluc^-$ $= Ca_3Gluc_2(OH)_4^0$ | - | - | | *(12.44)* 16.07 | - | *(12.59 [d])* 16.22 | **16.10** [e] |

Values reported for I = 0.1, 0.5 and 1 M extrapolated to 0 M ionic strength in this study using the WATEQ Debye Huckel equation (1); - : not reported; [a] values from Thoenen et al. (37); [b] value from Pallagi et al. (23); [c] notation $GlucOH^{2-}$ represents the two times deprotonated $C_6O_7H_{10}^{-2}$ as suggested in (24) ; [d] recalculated from reaction formulated with H+ (26) using a log $K_w$ of -13.62 at 1 M NaCl; [e] fitted in this study

The solubility of portlandite and the activity of $Ca^{2+}$ were fitted with a speciation model solved by the geochemical software PHREEQC version 3 (3.6.2-15100) (32) and the WATEQ4f database (33). The activity of the $Ca^{2+}$, $a_{Ca2+}$, and all other species was calculated according to $a_{Ca2+} = \gamma_{Ca2+} \cdot m_{Ca2+}$, where $\gamma_{Ca2+}$ is the activity coefficient and $m_{Ca2+}$ the molality in mol/kg $H_2O$. The activity coefficients were calculated with the WATEQ Debye Hückel equation:

$$\log \gamma_i = \frac{-A_y z_i^2 \sqrt{I}}{1 + B_y a_i \sqrt{I}} + b_y I \qquad (1)$$

where $z_i$ denotes the charge of species $i$, I is the effective molal ionic strength, while $a_i$ , the ion-size parameter, and $b_y$ are ion specific parameters and $A_y$ and $B_y$ are pressure- and temperature-dependent coefficients (33). This activity correction is applicable up to approximately 1 M of ionic strength (34).

**Table 2.** Complex formation constants, expressed in log K, between calcium and sorbitol (Sorb), mannitol (Man) and galactitol (Gal) at standard conditions (1 bar, 25°C) reported in literature and determined in the present study.

| | Kieboom (30) *(0.2-0.8M)* 0M | Haas (29) *(0.7M)* 0M | Kutus (25) *(1M)* 0M | This study 0M |
|---|---|---|---|---|
| $Ca^{2+}$+$Sorb^0$ = $CaSorb^{2+}$ | *(0.11)* 0.08 | *(-0.52)* -0.54 | *(0)* -0.06 | **0.10**[a] |
| $Ca^{2+}$ + $Sorb^0$ + $OH^-$ = $SorbCaOH^+$ | | | | **2.85**[a] |
| 2 $Ca^{2+}$ + 2$Sorb^0$ + 4$OH^-$ = $Ca_2Sorb_2(OH)_4^0$ | | | | **9.75**[a] |
| $Ca^{2+}$+ $Man^0$ = $CaMan^{2+}$ | *(-0.05)* -0.08 | *(-0.62)* -0.64 | *(-0.3)* -0.36 | **-0.36** |
| $Ca^{2+}$ + $Man^0$ + $OH^-$ = $CaManOH^+$ | | | | **2.65**[a] |
| 2 $Ca^{2+}$ + 2$Man^0$ + 4$OH^-$ = $Ca_2Man_2(OH)_4^0$ | | | | **9.65**[a] |
| $Ca^{2+}$ + $Gal^0$ = $CaGal^{2+}$ | | *(-0.51)* -0.53 | | **-0.53** |
| $Ca^{2+}$ + $Gal^0$ + $OH^-$ = $CaGalOH^+$ | | | | **2.80**[a] |
| 2 $Ca^{2+}$ + 2$Gal^0$ + 4$OH^-$ = $Ca_2Gal_2(OH)_4^0$ | | | | **9.29**[a] |

[a] Fitted in this study

The complex formation constants between $Ca^{2+}$ and hydroxide, gluconate, sorbitol, mannitol and galactitol reported in the literature and determined in the present study are detailed in Table 1 and Table 2. For gluconate, sorbitol and mannitol we used as starting values the complexes and associated constants derived from Pallagi and co-workers (23) (24) (25) (26). They were, where necessary, further refined to obtain a good visual fit between the measured and the modeled data. The following procedure to refine the complexation constants was employed. First, the potentiometric data measured at pH 11.3 were used to fit the constants for the $CaGluc^{+}$, $CaSorb^{2+}$, $CaMan^{2+}$, and $CaGal^{2+}$ complexes, which dominate at low pH values. Then, the titration data at higher pH values were used to fit the constants for the $CaGlucOH^{0}$, $CaSorbOH^{+}$, $CaManOH^{+}$, and $CaGalOH^{+}$ complexes. The formation of $CaGluc_2^{0}$, suggested by Pallagi et al. (23) and Kutus et al. (26), and of $Ca_2Gluc_2OH_4^{2-}$ and $Ca_3Gluc_2OH_4^{0}$complexes suggested by Kutus et al. (26), were also considered. Only traces of the $CaGluc_2^{0}$ complex were calculated to be present in our experiments (less than 1 % of the total Ca in any of the experiments) thus this complex was  considered but its constant not further refined. The presence of $Ca_3Gluc_2OH_4^{0}$ and $Ca_2Gluc_2OH_4^{2-}$ complex is not important at low calcium concentrations (i.e. the conditions used in this study for the potentiometric titration) but in the presence of portlandite. Thus their constants were refined using the solubility data of portlandite.

The $CaSorb^{2+}$, $CaMan^{2+}$, and $CaGal^{2+}$ complexes reported in the literature were not found to be in significant quantities in any of our experiments, as they were obtained in near neutral pH conditions. Furthermore, the reported values of the complex formation constants are rather scattered, maybe due to the different methods employed to determine them. We thus introduced in addition $CaHexitolOH^{+}$ complexes, which were found to give a very satisfactory description of our experimental data. As it will be described in the next section, the complexation of calcium by the hexitols is much weaker than by gluconate. The so-obtained complexation constant are compiled together with the literature values in Table 1 and 2.

## 3. Results and discussion

### 3.1 Gluconate

*3.1.1 Solubility experiment with portlandite*

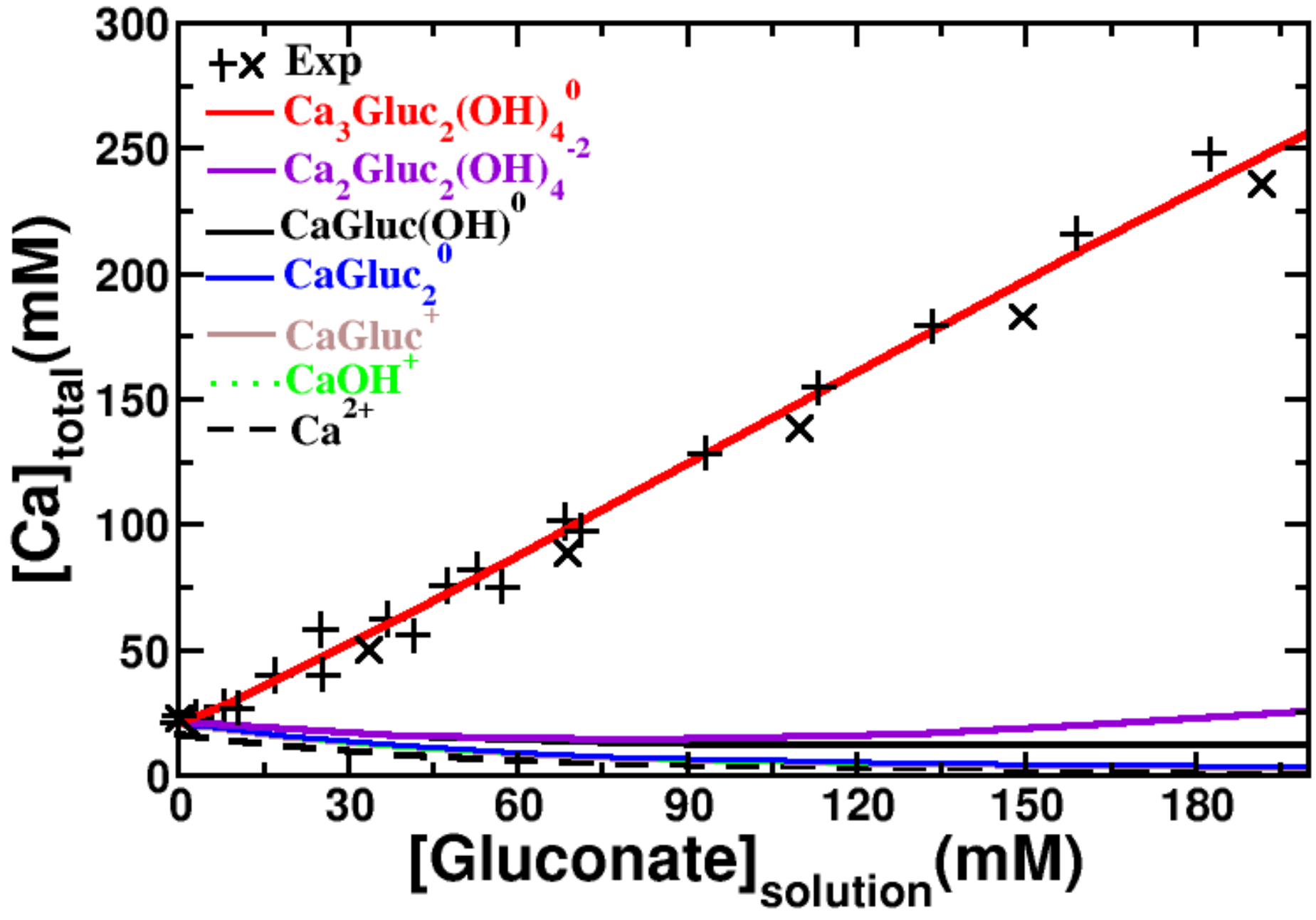

*Figure 4: Evolution of total calcium concentrations in equilibrium with portlandite (initial pH 12.6, final pH 12.8) as a function of the gluconate concentration. The crosses represent the total concentrations determined experimentally, while the solid lines represent the cumulative calcium complexes concentrations calculated using the data compiled in Table 1.*

The calcium concentrations in equilibrium with portlandite rise rapidly (Figure 4), when the gluconate concentration is increased. In the absence of gluconate, a calcium concentration of 21 mM was observed, which corresponds well with the expected solubility of portlandite of 21 mM at 23°C. The presence of gluconate increased the total measured calcium concentrations up to 101 mM Ca at a gluconate concentration of 68 mM, due to the formation of different Ca-gluconate complexes as shown in Figure 4. This strong increase of portlandite solubility in the presence of gluconate agrees well with observations reported by Nalet and Nonat (19). The measured increase of the total calcium concentrations ($[Ca]_{total} = [Ca^{2+}] + [CaOH^{+}] + [Ca\text{-organic}]$) is mainly due to the formation of $Ca_3Gluc_2(OH)_4^0$, $CaGlucOH^0$ and $Ca_2Gluc_2(OH)_4^{2-}$ in presence of gluconate, while our calculations indicate that the concentrations of $CaGluc^+$, and $CaGluc_2^0$ are negligible. The over-proportional increase of calcium (80 mM more calcium in solution in the presence of 68 mM gluconate) is consistent with the presence of the heteropolynuclear complex $Ca_3Gluc_2(OH)_4^0$, which contains more calcium than gluconate. At gluconate concentrations of above 20 mM the $Ca_3Gluc_2(OH)_4^0$ complex dominates Ca-speciation in the presence of portlandite as shown in Figure 4.

### *3.1.2 Ca-gluconate titration*

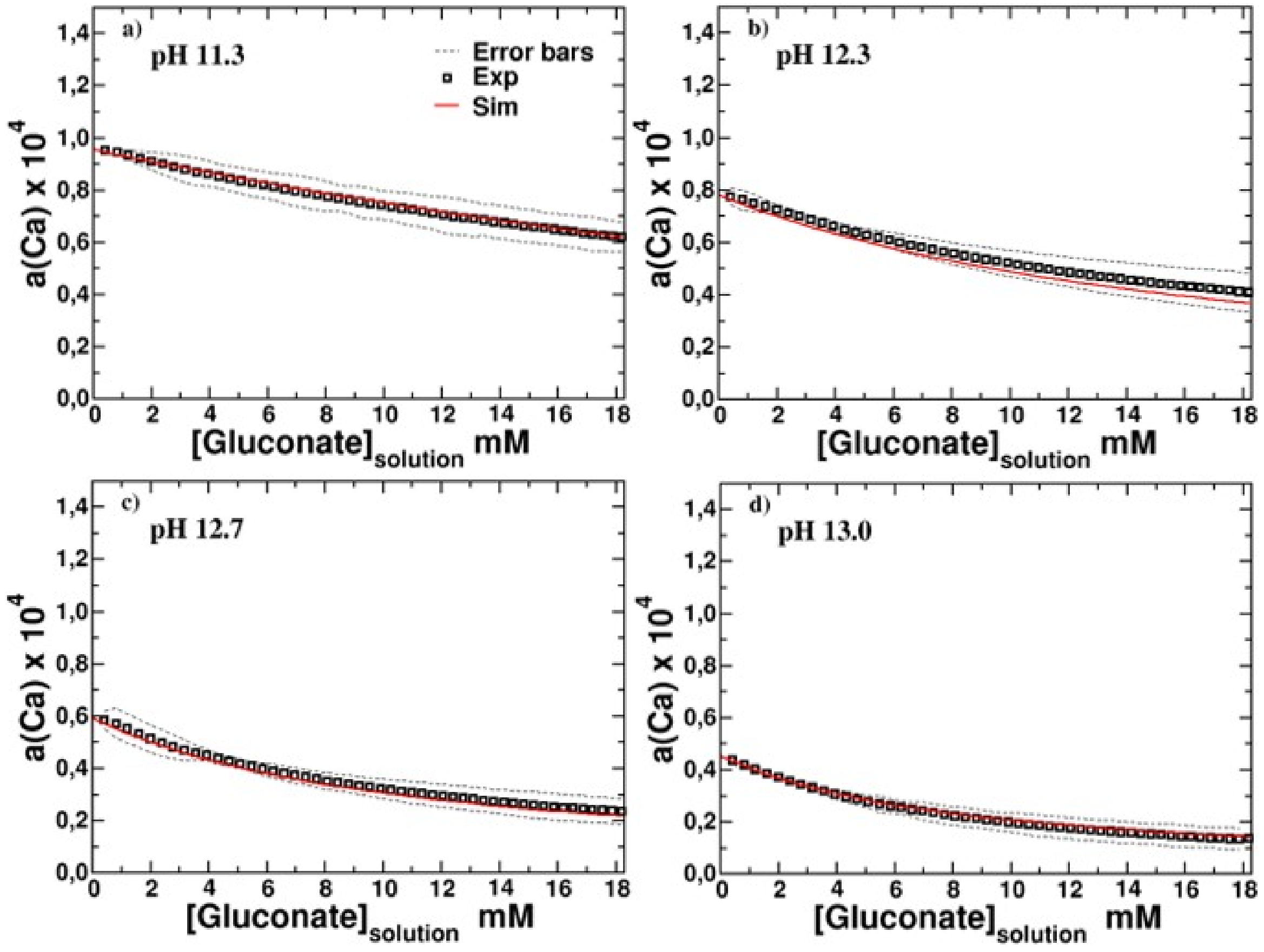

Figure 5: $Ca^{2+}$ activities, $a_{Ca}$, in a solution containing 0.25 mM $Ca(NO_3)_2$ and increasing amounts of 200 mM K-gluconate solution at pH a) 11.3, b) 12.3, c)12.7 and d) 13.0. The dots indicate the mean of three repetitions of the measurements and the dotted lines the observed standard deviations. The solid red lines show the modeled $a_{Ca}$ based on the data compiled in Table 1.

The measured changes of the $Ca^{2+}$ activity at various alkaline pH values upon the addition of potassium gluconate to a solution containing 0.25 mM calcium nitrate is shown in Figure 5. The drop of the measured $Ca^{2+}$ activity can be attributed i) to a minor extent to the dilution of the solution by the addition of the titrant solution (for the effect of adding solution without organics see SI) and ii) to the complexation of $Ca^{2+}$ with the added gluconate. As it can be seen in Fig. 5 a very good fit of the experimental data with our speciation model is obtained at all pH values. It can also be observed that the decrease in calcium activity is limited at pH 11.3 but more distinct at higher pH values, which points towards the importance of calcium-gluconate-hydroxide complexes, as further confirmed below.

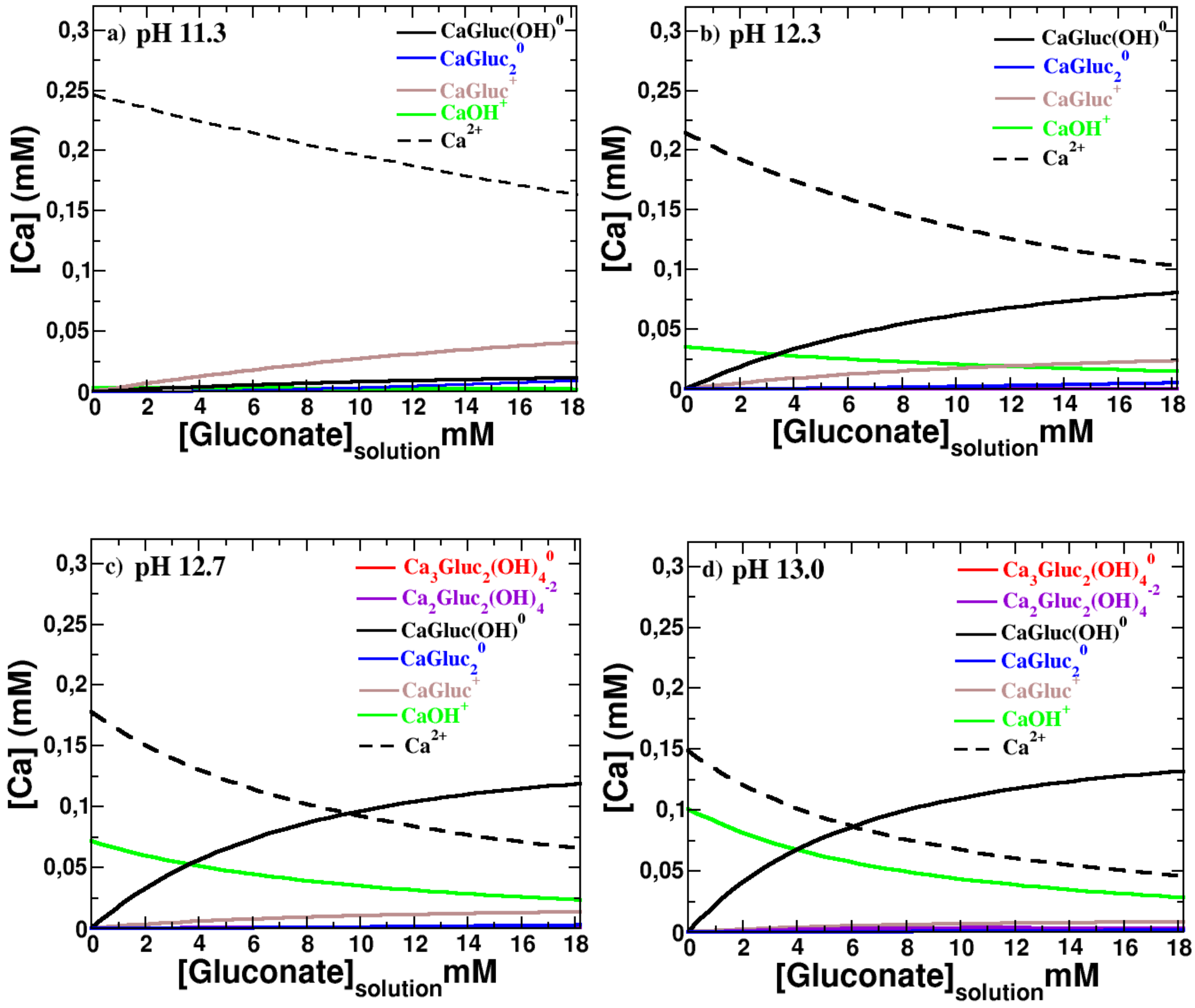

Figure 6: Calculated calcium concentrations (in mM) in a solution of 0.25 mM $Ca(NO_3)_2$ during the titration with 200 mM K-gluconate at pH a) 11.3, b) 12.3, c)12.7 and d) 13.0. The calculations are based on the thermodynamic data compiled in Table 1.

In Figure 6 the simulated change in the calcium speciation in the same conditions as in Figure 5 is given. In contrast in the solubility experiments, where high calcium concentrations (21 up to 100 mM Ca, see Figure 4) are present and the calcium complexation is dominated by the heteropolynuclear $Ca_3Gluc_2OH_4^0$ complex, $CaGluc^+$ and $CaGlucOH^0$ are the dominant complexes at the low Ca concentrations used in the titration experiments. For the titration at pH 11.3 mainly the formation of some $CaGluc^+$ is predicted, while at pH 13.0 the formation of $CaGlucOH^0$ is principally found, illustrating the strong influence of pH on the calcium speciation. In addition, it can be noted that calcium shows a strong preference for the heterogeneous complex $CaGlucOH^0$, whose concentration is 5 times higher in the presence of 18.8 mM gluconate than that of $CaOH^+$ at pH 13 (~100mM $OH^-$). Very low concentrations of the heteropolynuclear complexes, $Ca_2Gluc_2(OH)_4^{2-}$and $Ca_3Gluc_2(OH)_4^0$, were observed due to the relative low Ca (0.25 mM) and gluconate (18.8 mM) concentrations, at high pH ( 12.7 and 13.0 ).

## 3.2 Sorbitol

*3.2.1 Solubility experiments with portlandite*

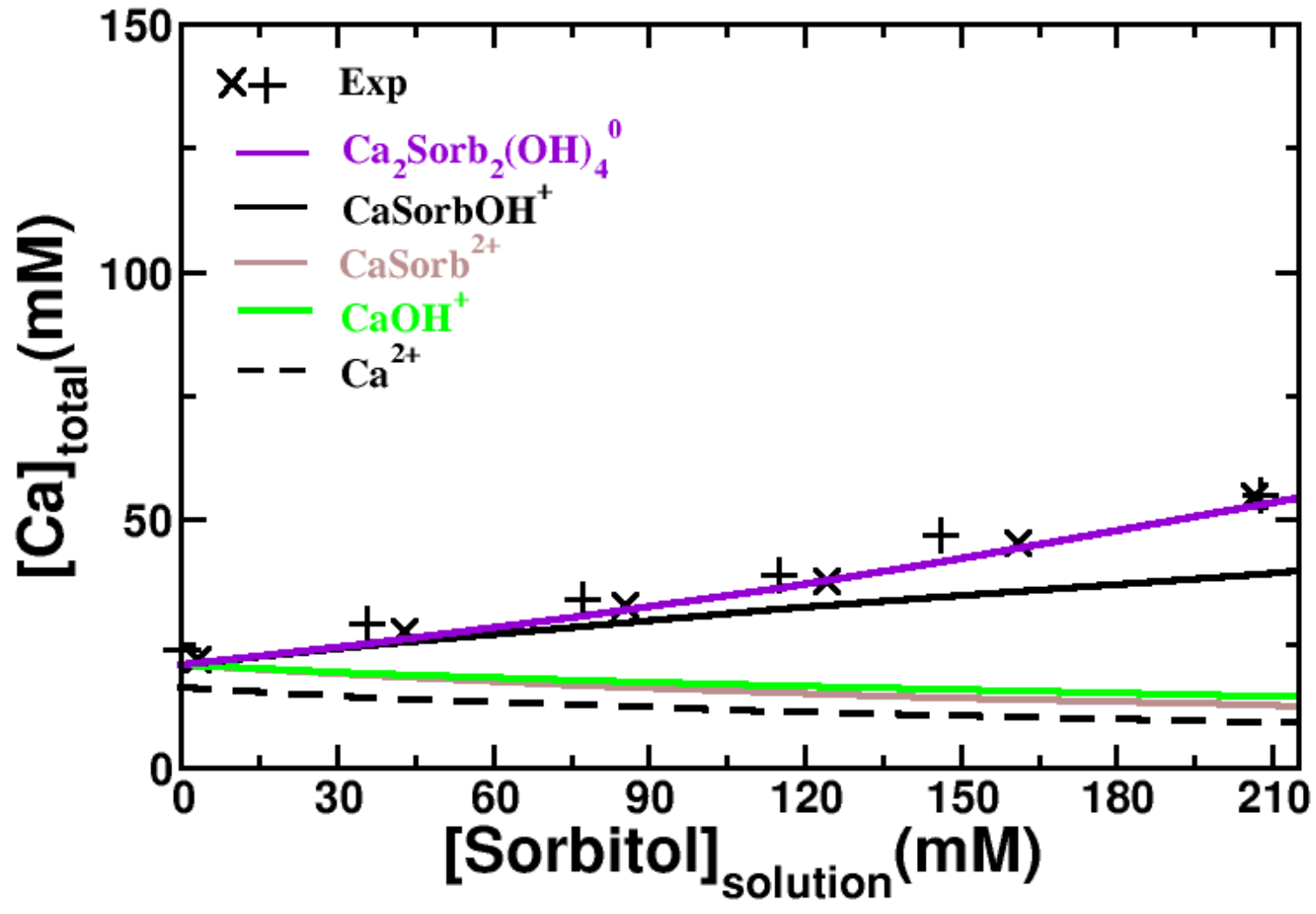

Figure 7: Evolution of the total calcium concentration in equilibrium with portlandite as a function of the sorbitol concentration. The crosses represent the total calcium concentrations determined experimentally, while the lines represent the calcium concentrations calculated using the data compiled in Table 2. The cumulative calcium concentrations due to $Ca^{2+}$ (black, dashed line), $CaOH^+$ (green, solid line), $CaSorb^{2+}$ (grey, solid line), $CaSorbOH^+$ (black, solid line) and $Ca_2Sorb_2(OH)_4^0$ (purple, solid line) are also plotted.

The equilibrium calcium concentration as obtained in the solubility experiments of portlandite in the presence of sorbitol is given in Figure 7. The equilibrium calcium concentration is observed to increase moderately with that of sorbitol, from 21 mM to 55 mM when sorbitol is increased up to 211 mM. The increase is much weaker than the one due to gluconate (Figure 4), indicating a weaker complex formation between $Ca^{2+}$ and sorbitol. The calculations show that the observed increase can be mainly explained by the formation of a $CaSorbOH^+$ complex, while the concentration of $CaSorb^{2+}$ is found to be negligible. No clear indication for the formation of polynuclear complexes is found, although the underestimation of the total calcium concentration at very high sorbitol concentrations could point towards the formation of such complex. At sorbitol concentrations of 100 mM and above, the $CaSorbOH^{+}$ complex dominates Ca-speciation (Figure 7).

*3.2.2 Ca-sorbitol titration*

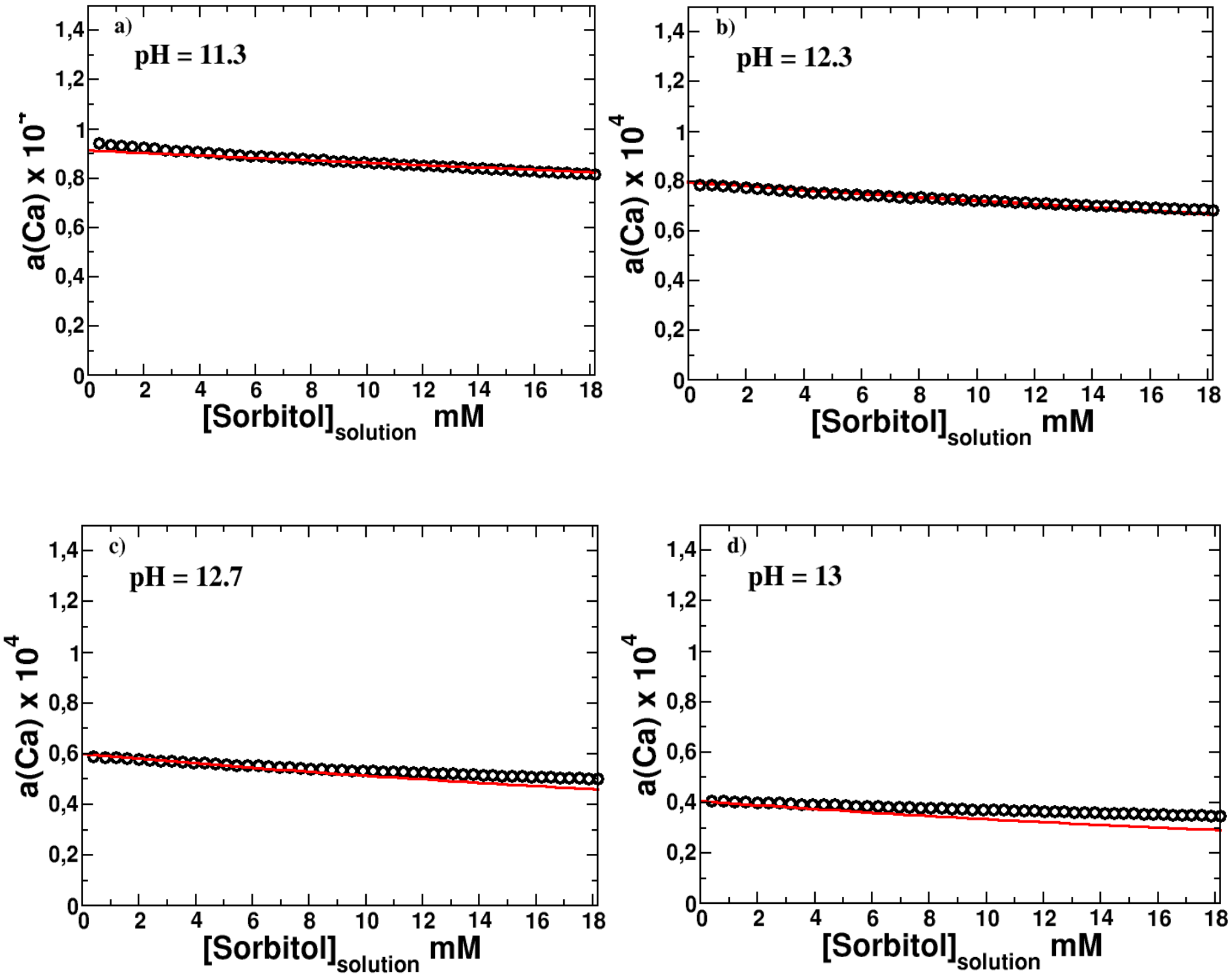

Figure 8: $Ca^{2+}$ activities, $a_{Ca2+}$, in a solution containing 0.25 mM $Ca(NO_3)_2$ and increasing amounts of 0.2 M sorbitol solution at pH a) 11.3, b) 12.3, c) 12.7 and d) 13.0. The solid red lines show the modeled $a_{Ca2+}$ based on the data compiled in Table 2.

The change in the activity of $Ca^{2+}$ at different pH values upon the addition of sorbitol as measured by potentiometric titration of a diluted calcium nitrate solution (0.25 mM) is shown in Figure 8. In agreement with the solubility experiments (Fig.7), the drop of the $Ca^{2+}$ activity is weaker than the one observed with gluconate and more distinct at high pH values as explained by the formation of $CaSorbOH^{+}$ complex. This is illustrated in Figure 9, which provides the detailed calculated speciation of calcium. As expected, $CaSorbOH^{+}$ is prevalent at pH 13 and hardly visible at pH 11.3.

In line with the solubility experiments, the $CaSorb^{2+}$ complex is negligible in these alkaline conditions. The overall good agreement obtained between modeled and experimentally observed decrease of the $Ca^{2+}$ activities clearly shows that no or only very little polynuclear Ca-sorbitol complexes are present.

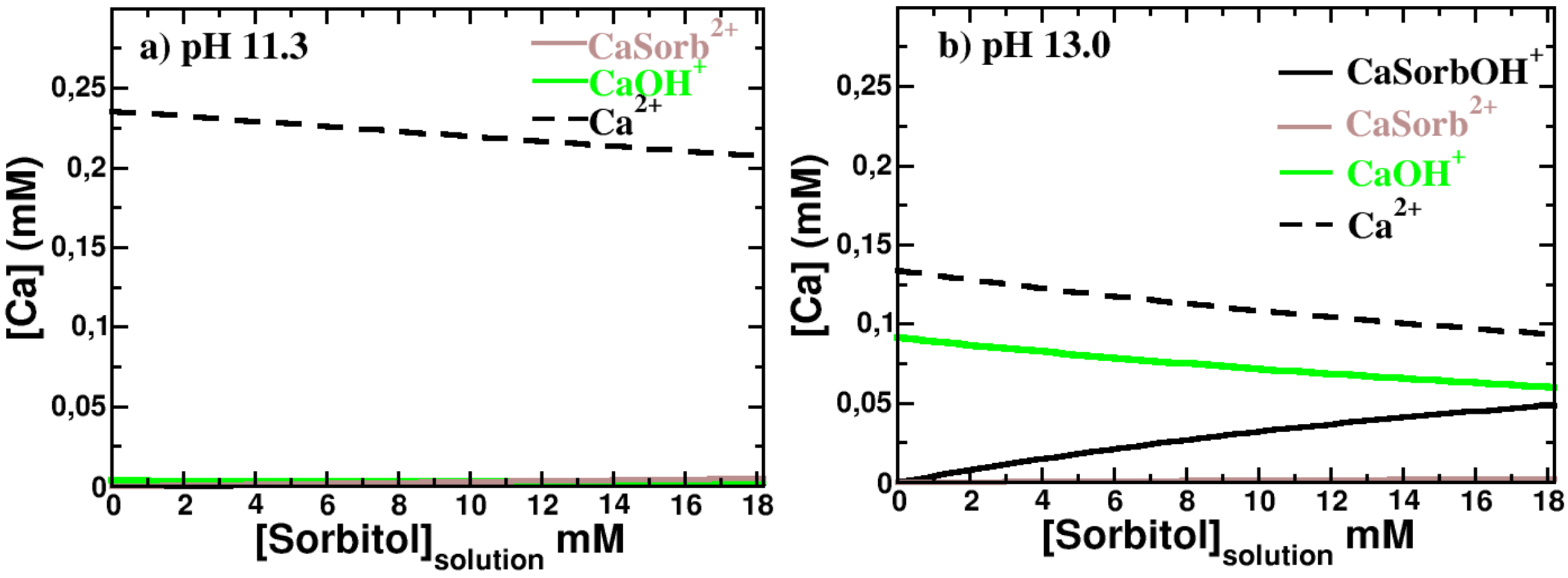

Figure 9: Calcium concentrations (in mM) in a solution of 0.25 mM $Ca(NO_3)_2$ during the titration with sorbitol at a) pH 11.3 and b) pH 13.0 calculated based on the thermodynamic data compiled in Table 2.

## 3.3 Mannitol and galactitol

### *3.3.1 Solubility experiments with portlandite*

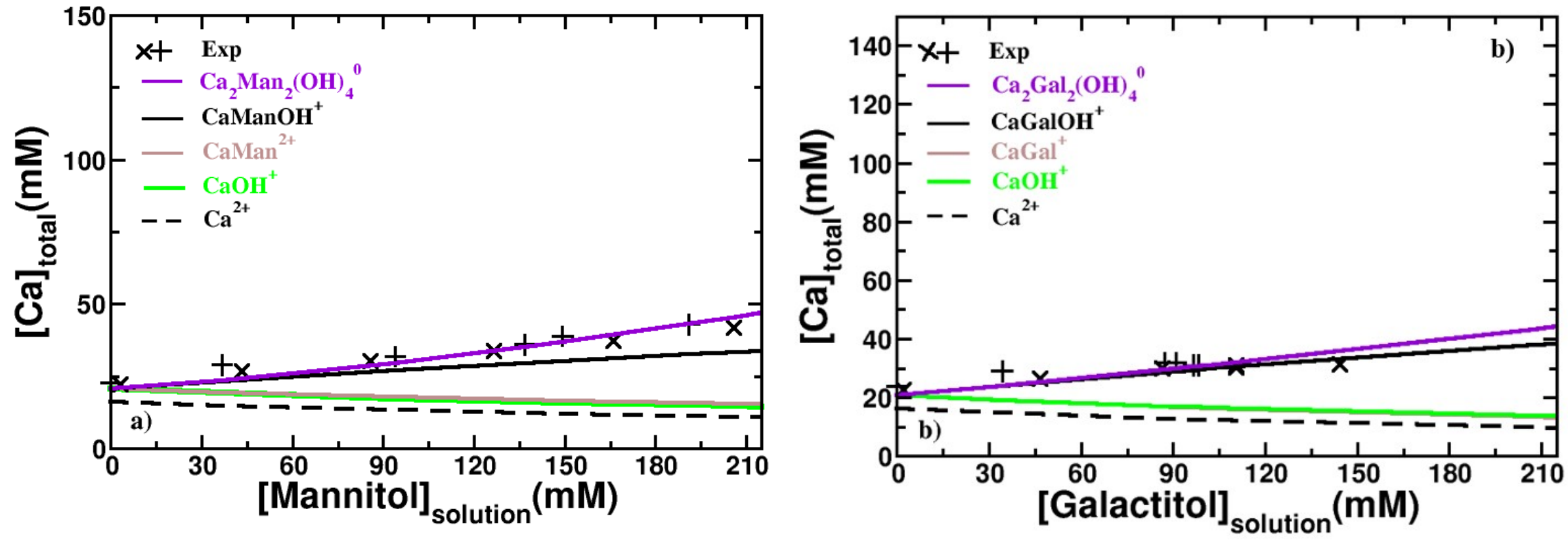

Figure 10: Evolution of calcium concentrations in equilibrium with portlandite at a pH value of 12.6 as a function of a) mannitol and b) galactitol concentration. The crosses represent the total calcium concentrations determined experimentally, while the lines represent the calcium concentrations calculated using the thermodynamic data compiled in Table 2. The cumulative calcium concentrations of $Ca^{2+}$ (black, dashed line), $CaOH^{+}$(green, solid line), $CaMan^{2+}$ or $CaGal^{2+}$ (grey, solid line), $Ca_2Man_2(OH)_4^0$ or $Ca_2Gal_2(OH)_4^0$ (purple, solid line) and $CaManOH^{+}$ or $CaGalOH^{+}$ (black, solid line) are also plotted.

The increase in the equilibrium calcium concentration in the solubility experiments of portlandite in the presence of mannitol and galactitol, given in Figure 10, is comparable to that observed with sorbitol, although somewhat weaker compared with Figure 7. As for sorbitol, the observed increase of the Ca concentration can be principally explained by the formation of $CaManOH^{+}$ and $CaGalOH^{+}$ complexes, while the concentrations of $CaMan^{2+}$ and $CaGal^{2+}$ are negligible. Only at mannitol and galactitol concentrations well above 100 mM, $CaManOH^{+}$ and $CaGalOH^{+}$ complexes dominate the speciation of calcium (Figure 10).

*3.3.2 Ca-mannitol and Ca-galactitol titration*

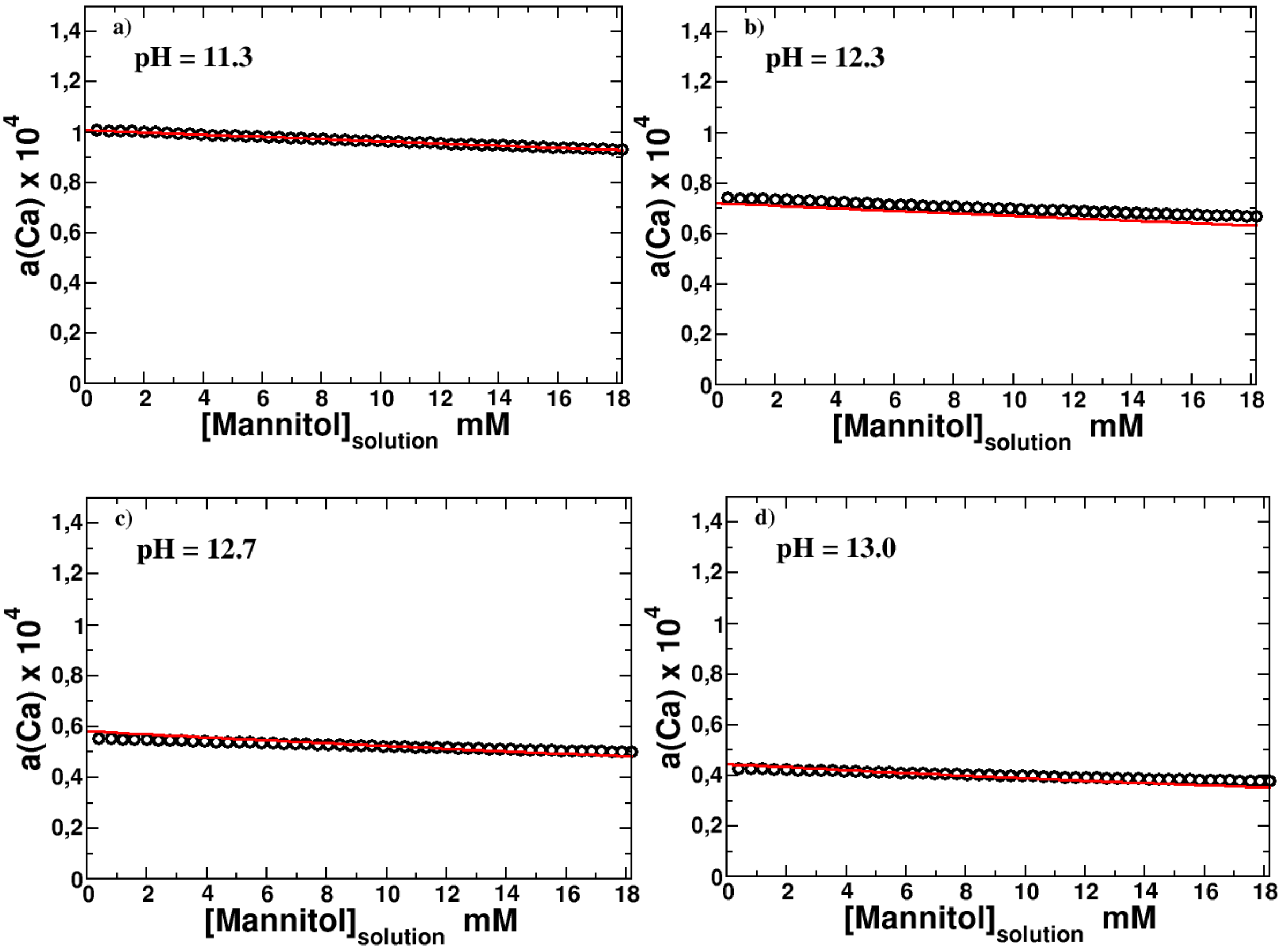

Figure 11: $Ca^{2+}$ activities, $a_{Ca2+}$, in a solution containing 0.25 mM $Ca(NO_3)_2$ and increasing amounts of 0.2 M mannitol solution at pH a) 11.3, b) 12.3, c)12.7 and d) 13.0. The experimental points are shown by the empty circles. The solid red lines give the modeled $a_{Ca2+}$, based on the data compiled in Table 2, for comparison.

The change in the simulated and measured activity of $Ca^{2+}$ and pH upon the addition of 0 to 18 mM mannitol to a 0.25 mM calcium nitrate at pH 11.3, 12.3, 12.7 and 13.0 are shown in Figure 11. The data for galactitol are similar and provided in the supplementary information. In agreement with the solubility experiments of portlandite (Figure 10), the decrease in the measured $a_{Ca2+}$ is less pronounced than in the case of sorbitol and is mainly explained by the formation $CaManOH^+$ and $CaGalOH^{+,}$ at higher pH values. The strong preference of calcium for the heterogeneous complex with hydroxide ($CaManOH^+$ and $CaGalOH^+$) is further illustrated in Figure 12.

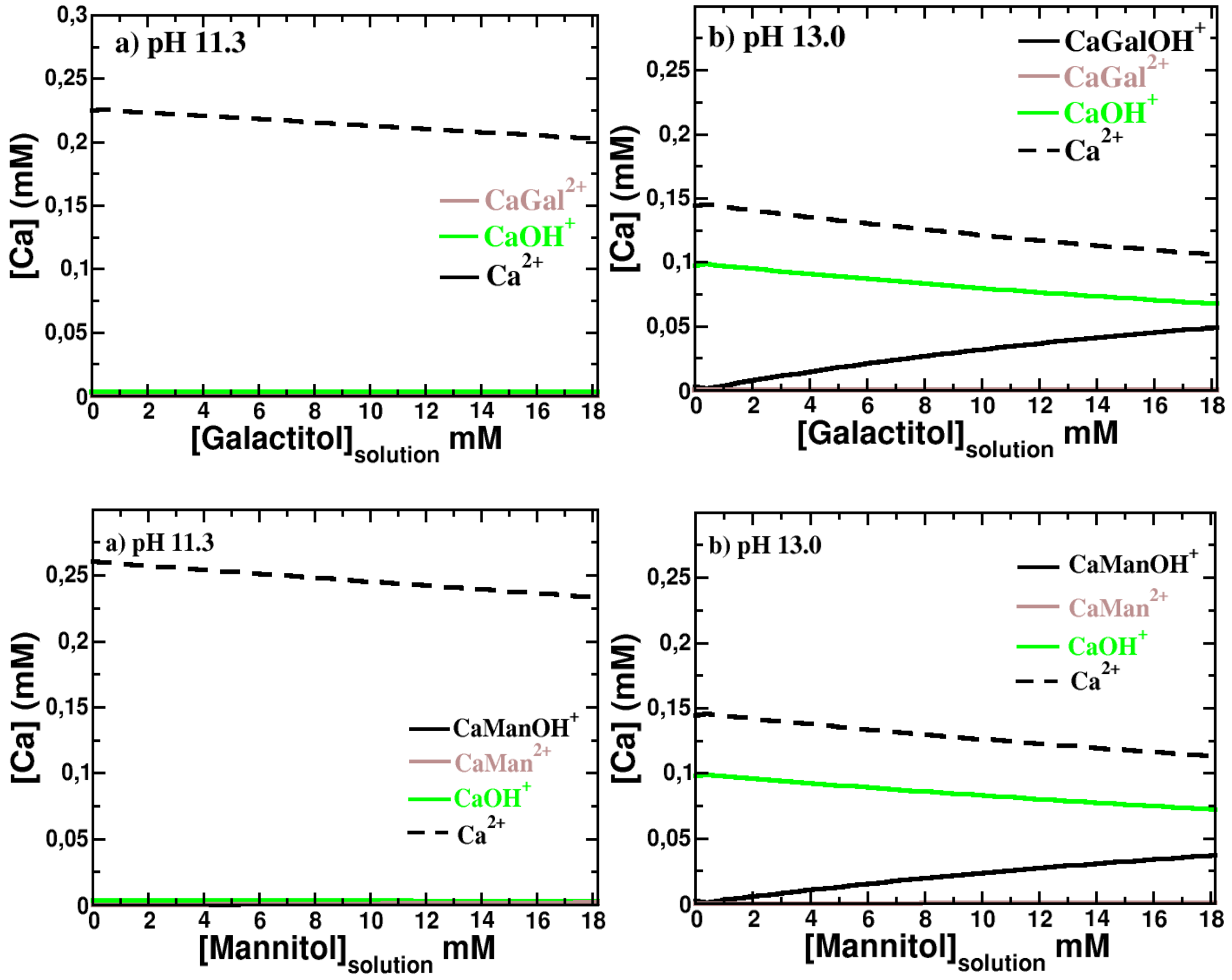

Figure 12: Simulated speciation of calcium in a solution of 0.25 mM $Ca(NO_3)_2$ during the titration with above a) galactitol at pH11.3 and b) galactitol at pH 13.0, and below, a) mannitol at pH 11.3 and b) mannitol at pH 13.0. The calculations are based on the thermodynamic data compiled in Table 2.

## 3.4 Effect of complexation on calcium speciation

Calcium has been observed to form a number of different complexes with gluconate and hydroxide. Under conditions relevant for the early-age pore solution of cements (10-40 mM Ca, pH 12.5 -13.5) mainly the $CaGlucOH^0$, $Ca_3Gluc_2(OH)_4^0$ and $Ca_2Gluc_2(OH)_4^{-2}$ complexes are of importance as illustrated in Figure 13b. The importance of $CaGlucOH^0$, $Ca_3Gluc_2(OH)_4^0$ and $Ca_2Gluc_2(OH)_4^{-2}$ complexes at pH values above 12.5 result in much lower concentrations of free $Ca^{2+}$ than in the absence of gluconate. The effect can be expected to be even stronger at later hydration times, where calcium concentrations drop to a few mM, while the concentrations of small organic molecules in the pore solutions tend to remain high. This leads to a stabilization of $CaGlucOH^0$ as shown in Figure 13a and lower $Ca^{2+}$ concentrations.

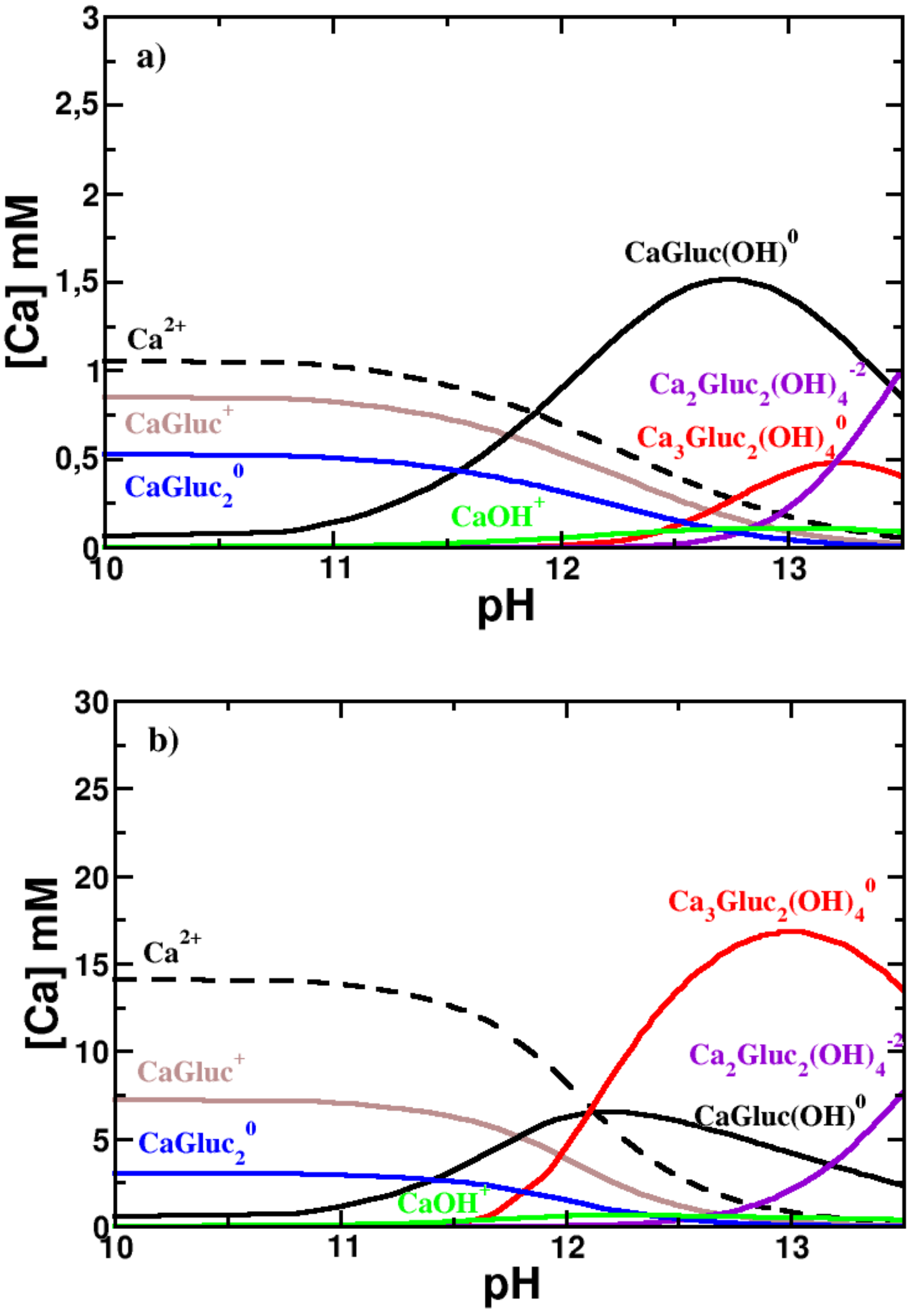

Figure 13: Calcium distribution (expressed as mM Ca) in a solution containing a) 2.5 mM of Ca , b) 25 mM of Ca and 50 mM of gluconate in the pH range 10 to 13.5.

The strong complexation of calcium can be expected to retard portlandite and C-S-H precipitation during cement hydration. Gluconate sorbs also strongly on calcium at the surface of $C_3S$, portlandite (see Supplementary Information) and C-S-H (18) (19), which will also strongly influence their dissolution and formation rate.

Calcium forms only relatively weak complexes with the three hexitols investigated, in the order sorbitol > mannitol > galactitol. Thermodynamic modelling indicates a non negligeable Ca-complexation in conditions relevant for the pore solution of cements (10-40 mM Ca, pH 12.5 -13.5) as illustrated for sorbitol in Figure 14.

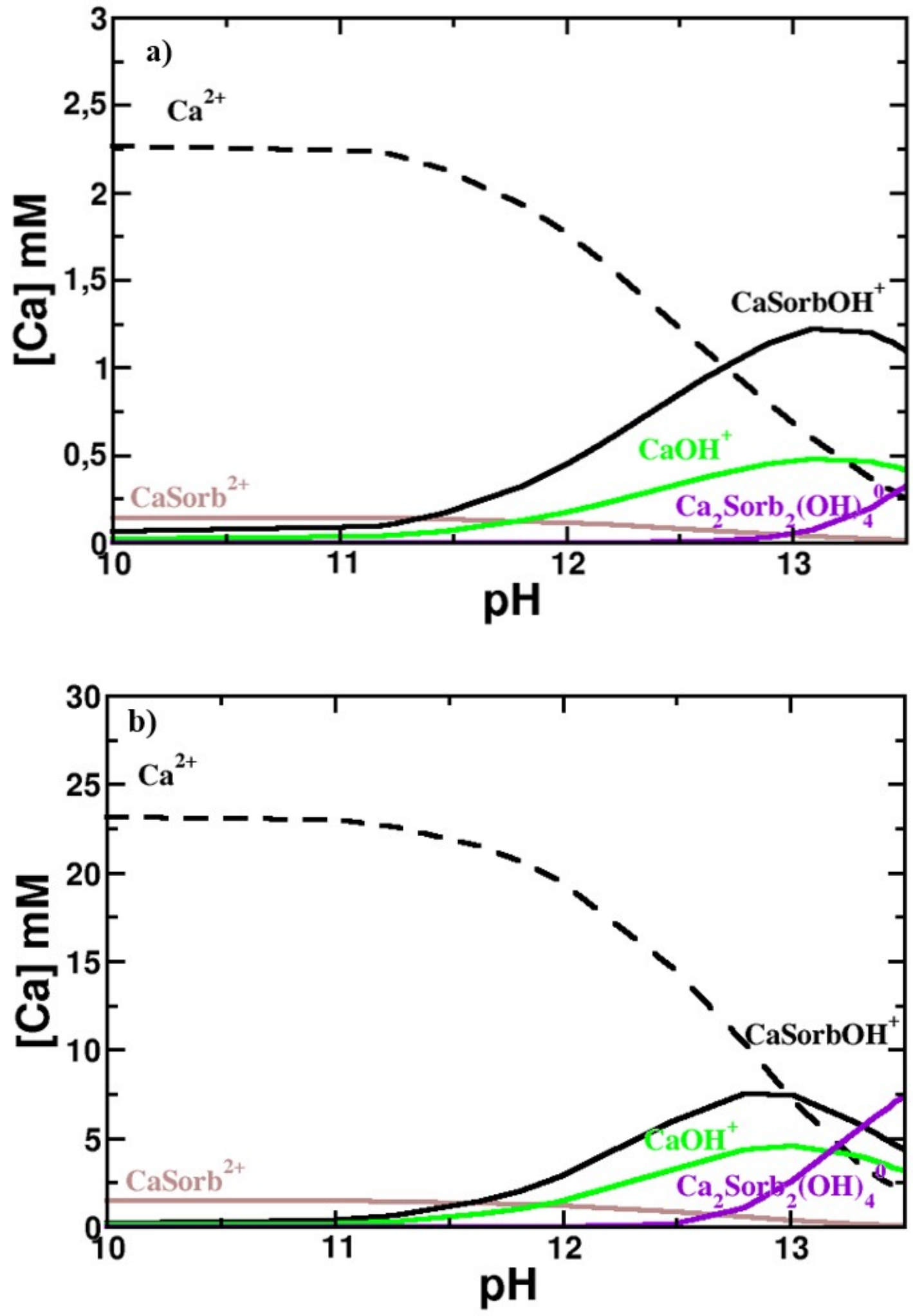

Figure 14: Calcium distribution (expressed as mM Ca) in a solution containing a) 2.5 mM of Ca, b) 25 mM of Ca and50 mM of sorbitol in the pH range 10 to 13.5.

The observed tendency of calcium to form complexes with organics follows the order gluconate >> sorbitol > mannitol > galactitol, which corresponds well with the tendency to sorb on portlandite (see Supplementary informations) and C-S-H (18) (19): gluconate >> sorbitol > mannitol, but only partially with their tendency to retard the $C_3S$ hydration reported in Nalet and Nonat (15): gluconate >> sorbitol > galactitol > mannitol. The reason for the different sequence of galactitol and mannitol on $C_3S$ hydration is presently not clear. The charged gluconate, which complexed strongly with calcium in solution had also the biggest retarding effect on $C_3S$ hydration.

## 4. Conclusions

The complexation of $Ca^{2+}$ with gluconate, D-sorbitol, D-mannitol and D-galactitol has been studied via portlandite solubility measurement and titration experiments at low ionic strength (0.1 M $KNO_3$).

For gluconate, the multinuclear complexes already described in the literature allowed us to describe the experimental data, after some further refinement of the complexation constants. At a pH of 12.5 and in the presence of portlandite the heteropolynuclear complex $Ca_3Gluc_2(OH)_4^0$ dominates the Ca-speciation, while at lower calcium concentrations $CaGluc^+$ (below pH 12) and $CaGlucOH^0$ (above pH 12) are the main complexes formed. This relative strong complex formation between calcium and gluconate lowers concentrations of free $Ca^{2+}$, which could contribute to a retardation of portlandite and C-S-H precipitation during cement hydration. The strong tendency of gluconate to form complexes with Ca reported here is consistent with the significant sorption of gluconate on Ca on the surface of C-S-H and portlandite reported (19) .

Sorbitol makes weaker complexes with calcium as observed both from portlandite solubility measurements and titration results. Under all conditions studied, the predominant sorbitol complex is the ternary $CaSorbOH^+$ complex, while the $CaSorb^{2+}$ complex formed in negligible amounts only. In all cases studied, $CaSorbOH^+$ complex had limited effect on calcium speciation below a pH of 12, but can dominate the calcium speciation at pH 12.5 and above, at higher sorbitol concentrations.

Similar observations have been made for D-mannitol and D-galactitol, which show an even weaker tendency than sorbitol to form calcium complexes. Also, for D-mannitol and D-galactitol only the ternary $CaManOH^+$ and $CaGalOH^+$ complexes are relevant, and they are expected to form mainly above pH 12.5 and at high mannitol and galactitol concentrations.

The observed tendency of calcium to form complexes follows the order gluconate >> sorbitol > mannitol > galactitol, which corresponds well with the tendency to sorb on portlandite and C-S-H: gluconate >> sorbitol > mannitol (19) but only partially with their tendency to retard the $C_3S$ hydration reported in (15): gluconate >> sorbitol > galactitol > mannitol.

**Acknowledgements**

The financial support from Nanocem (core project 15) is thankfully acknowledged. We also would like to thank the representatives of the industrial partners: L. Pegado, J.H. Cheung, V. Kocaba, P. Juilland, and M. Mosquet for many helpful discussions and their interest in this project. We sincerely thank L. Brunetti, S. El Housseini and D. Nguyen for their help in the laboratory work. The use of the analytical platform of ISTerre, with the help of D. Tisserand, S. Bureau and S. Campillo, is acknowledged.